\documentclass[10pt, pre, amsmath, twocolumn, showpacs, superscriptaddress]{revtex4-1}
\usepackage{amsmath}
\usepackage{amssymb}
\usepackage{mathtools}

\usepackage{graphicx}

\usepackage{color}
\definecolor{dgreen}{rgb}{0,0.7,0}

\usepackage{caption}
\usepackage{subcaption}

\newcommand{\dd}{\mathrm{d}}
\newcommand{\ee}{\mathrm{e}}

\newcommand{\cO}{\mathcal{O}}

\newcommand{\pp}{p'}
\newcommand{\qp}{q'}
\newcommand{\rp}{r'}
\newcommand{\deltap}{\delta'}
\newcommand{\vtr}{{v_\text{tr}}}
\newcommand{\vtrh}{v_\text{tr,H}}
\newcommand{\vtre}{v_\text{tr,E}}
\newcommand{\rhob}{\overline{\rho}}
\newcommand{\muee}{\mu_{\text{E},\text{E}}}
\newcommand{\Jb}{\mathcal{J}_\mathrm{B}}

\begin{document}

\title{Driven tracer with absolute negative mobility}

 \author{J. Cividini}
 \email{julien.cividini@weizmann.ac.il}
 \affiliation{Department of Physics of Complex Systems,
 Weizmann Institute of Science\\
 Rehovot, Israel 7610001}

 \author{D. Mukamel}
 \email{david.mukamel@weizmann.ac.il}
 \affiliation{Department of Physics of Complex Systems,
 Weizmann Institute of Science\\
 Rehovot, Israel 7610001}

\author{H.A. Posch}
\email{Harald.Posch@univie.ac.at}
\affiliation{Computational Physics Group, Faculty of Physics, 
 Universit\"at Wien,
 Boltzmanngasse 5, 1090 Vienna, Austria}

\date{\today}

\begin{abstract}
Instances of negative mobility, where a system responds to a perturbation in a way opposite to naive expectation, have been studied theoretically and experimentally in numerous nonequilibrium systems. In this work we show that Absolute Negative Mobility (ANM), whereby current is produced in a direction opposite to the drive, can occur around \emph{equilibrium states}. This is demonstrated with a simple one-dimensional lattice model with a driven tracer. We derive analytical predictions in the linear response regime and elucidate the mechanism leading to ANM by studying the high-density limit. We also study numerically a model of hard Brownian disks in a narrow planar channel, for which the lattice model can be viewed as a toy model. We find that the model exhibits Negative Differential Mobility (NDM), but no ANM.
\end{abstract}

\maketitle


\section{Introduction}

Take a system in an equilibrium or nonequilibrium steady state and apply a small drive to it.  
We expect the system to move in the direction of the drive, and increasingly so with stronger drives. However, various setups have been found where intuition is violated and the response is more surprising. Among the simplest manifestations of this behavior are Negative Differential Mobility (NDM), where the response coefficient depends on the applied perturbation in a nonmonotonic way, and Absolute Negative Mobility (ANM), where the sign of the response coefficient is opposite to what would intuitively be expected.

Theoretical examples of NDM include uniformly driven systems~\cite{bottger_b1982, hopfel_s_g1986, vrhovac_p1996, benenti2009, turci_p_s2012, reichhardt_r2017} or single driven tracers in quiescent media~\cite{cecchi_m1996, slater_g_n1997, zia_p_m2002, perondi2005, kostur2006, jack2008, sellitto2008, leitmann_f2013, baerts2013, basu_m2014, benichou2014, baiesi_s_v2015, benichou2016, sarracino2016, cecconi2017, burlatsky1992, burlatsky1996, landim_o_v1998, benichou1999, deconinck_o_m1997, benichou2001, benichou2000b, benichou2015, brummelhuis_h1989, illien2013, illien2015, benichou2013a,benichou2013b,benichou2013c, benichou2016, demery_d2010a, demery_d2011, cividini2016a, cividini2016b, cividini_m_p2017}. Also condensed matter experiments have been performed~\cite{conwell1970, nava1976, bottger_b1982, stanton_b_w1986, lei_h_c1991}.
The appearance of NDM mostly relies on trapping mechanisms that can be implemented through \textit{e.g.} complicated potentials~\cite{stanton_b_w1986, cecchi_m1996, kostur2006, sarracino2016, cecconi2017} or impurities, either present by definition~\cite{zia_p_m2002, perondi2005, jack2008, sellitto2008, leitmann_f2013, baerts2013, baiesi_s_v2015} or effectively created by a slow relaxation of the surrounding medium~\cite{turci_p_s2012, basu_m2014, benichou2014, benichou2016}. A pedagogical explanation for NDM is given in Ref.~\cite{zia_p_m2002}, and a modified Green-Kubo formula that accounts for NDM has been proposed in Ref.~\cite{baerts2013}.

Absolute Negative Mobility has been observed in a variety of setups. 
Typically, one does not expect ANM to take place when the unperturbed system is in equilibrium, since, as has been argued, this would constitute a violation of the Second Law of Thermodynamics~\cite{eichhorn_r_h2002, eichhorn_r_h2002b, cleuren_v2002}. Thus, previous studies demonstrating ANM considered a driving field which acts on nonequilibrium steady states. These include systems with a periodic~\cite{keay1995, ignatov1995, hartmann_g_h1997, aguado_p1997, cannon2000, machura2007, eichhorn2010, spiechowicz_l_h2013, slapik_l_s2018, hanggi_m2009, speer_e_r2007} or a random~\cite{goychuk_p_m1998, haljas2004, spiechowicz_h_l2014b, spiechowicz_l_m2016, hanggi_m2009} drive, random walkers~\cite{cleuren_v2002, eichhorn_r_h2002, eichhorn_r_h2002b, sarracino2016}, strong interactions and noise in spatially periodic potentials~\cite{reimann1999, reimann_b_k1999, buceta2000, cleuren_v2001, mangioni_d_w2001}, and others~\cite{rozenberg_l_r1988, reguera_r_p2000, ghosh2014, dotsenko_m_o2017}. A different setup where ANM has been found both experimentally and theoretically involves quantum mechanical effects such as absolute negative conductivity for semiconductors, where negative conductivity is associated with a negative effective mass of the carriers (either electrons or holes)~\cite{kromer1958, mattis_s1959, cannon2000}, or interactions between light and matter~\cite{aronov_s1975, gershenzon_f1986, gershenzon_f1988, dakhnovskii_m1995, aguado_p1997}.
 
In the present work we focus on cases, where the unperturbed system is in \emph{thermal equilibrium}, and demonstrate that, depending on the drive mechanism, ANM can take place in such systems. This is done in the context of a model of a tracer moving on a discrete ring populated by neutral particles, which obey a Simple Symmetric Exclusion Process (SSEP) -type dynamics. We show that, when a driving force is applied to the tracer, it moves in a direction opposite to the
drive. We then consider a continuum analogue of the model by studying the Langevin-type motion of a tracer particle in a narrow channel of a  gas of hard disks. Here we find that the model exhibits NDM but not ANM.

In the discrete ring model introduced in the present work, the dynamics of the tracer is such that it can move by two different processes, either by hopping towards a neighboring vacant site or by exchanging its position with close bath particles.
The exchange move requires enough 'free volume' in the vicinity, which places restrictions  on the dynamics and is reminiscent of kinetically-constrained models~\cite{ritort_s2003, jack2008, sellitto2008, turci_p_s2012}.
The system is studied analytically within a mean-field approximation and numerically, and the existence of ANM is demonstrated. The model may be considered as a kind of  toy model for hard disks performing Brownian motion dynamics in a narrow planar channel. Therefore, we carried out Langevin-type  simulation studies of this hard disks model where NDM but no ANM has been found. But we believe that some simple variants of this model, which have not yet been tested, could exhibit ANM for selected sets of parameters.

The paper is organized as follows: in Section~\ref{section:ssep} we study the lattice model analytically and numerically. In Section~\ref{section:hd} we introduce the model of hard disks moving in a narrow channel and present the results of molecular dynamics simulations. In Section~\ref{section:ccl} we conclude with a discussion of the results.

\section{Hard-core particles on a lattice}
\label{section:ssep}

\subsection{Definition}

Consider a set of $N$ bath particles and a tracer particle occupying $L$ sites of a ring of length $L$ while satisfying the simple exclusion constraint. Time is continuous and each transition occurs with a probability $R \dd t$ during each infinitesimal time step $\dd t$, where $R$ is the rate of the transition. The bath particles are regular SSEP particles and can hop towards the site directly to their left or to their right, each with constant rate $1$, under the condition that the target site is vacant. Their average density is denoted by $\rhob = \frac{N}{L-1}$.

\begin{figure}[!ht]
	\begin{center}
		\includegraphics[width=0.4\textwidth]{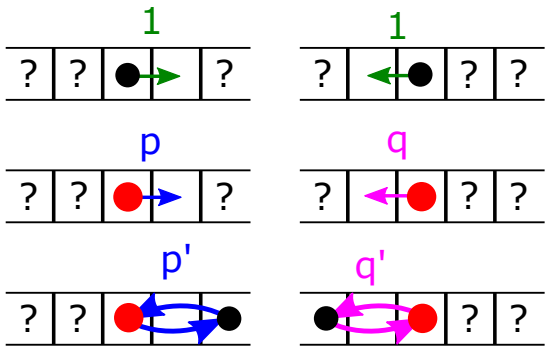} 
	\end{center}
	\caption{\small Allowed transitions. Bath particles (black) hop to the right and to the left with constant rate $1$ each (top row), the tracer (red) hops to the right with rate $p$ and to the left with rate $q$ (middle row) and can exchange its position with a particle two sites away with rates $\pp$ and $\qp$ if the intermediate site is empty (bottom row).}  
	\label{fig:system}
\end{figure}

The tracer is different from the bath particles in two ways. First, it hops to the right and to the left with different rates that we call $p$ and $q$, respectively. Second, it can exchange its position with a bath particle two sites away under the condition that the site between the tracer and the bath particle is vacant. This process takes place with rate $\pp$ to the right and $\qp$ to the left. The condition that the intermediate site has to be empty mimics the fact that, in more realistic systems of \textit{e.g.} particles moving in a narrow channel, overtakes are easier when particles have more space.  
The allowed transitions are summarized in Fig.~\ref{fig:system}. Such a system can be easily simulated using the Monte Carlo algorithm.

At large times we expect the tracer to have a finite velocity and the bath to reach a stationary nonequilibrium state in the frame of the tracer. We define the occupation variables in the frame of the tracer by $\tau_l = 0$ or $1$ for $l=1,\ldots,L-1$ and use $\langle \ldots \rangle$ for ensemble averages. The tracer occupies site $l=0$.
The average velocity of the tracer is given by
\begin{eqnarray}
\label{eq:vtrdef}
\vtr &=&  p (1-\langle \tau_1 \rangle) -q (1- \langle \tau_{L-1} \rangle) \nonumber \\
&&+ 2 \pp \langle (1-\tau_1) \tau_2 \rangle - 2 \qp \langle (1-\tau_{L-1}) \tau_{L-2}\rangle.
\end{eqnarray}
The first term accounts for hops of the tracer one step to the right: the transition is allowed if site $l=1$ is empty, contributing a factor $1-\tau_1$, and then occurs with rate $p$. The third term accounts for exchanges to the right: this transition is allowed if site $l=1$ is empty and site $l=2$ is occupied (factor $\tau_2(1-\tau_1)$). It occurs with rate $\pp$, and the tracer moves two steps to the right (factor $2$). The second and fourth terms are hops and exchanges to the left, respectively. An ensemble average of the whole expression is taken.
We also define the densities $\rho_l = \langle \tau_l \rangle$ for $l=1,\ldots,L-1$.

A configuration of the system is entirely specified by the $\{\tau_l\}_{l=1,\ldots,N-1}$, supplemented with the position of the tracer in the lab frame. In the case $p=q$ and $\pp = \qp$ it is clear that the rate of every allowed transition between two states of the system is equal to the rate of the inverse transition. This implies that detailed balance is satisfied and that the stationary distribution is flat.

Interesting phenomena happen when a drive is applied to the system, namely $q \neq p$ or $\qp \neq \pp$. We now present numerical results for the tracer velocity and the density profile before showing how they can be understood analytically.

\begin{figure} 
	\begin{center}
		\includegraphics[width=0.45\textwidth]{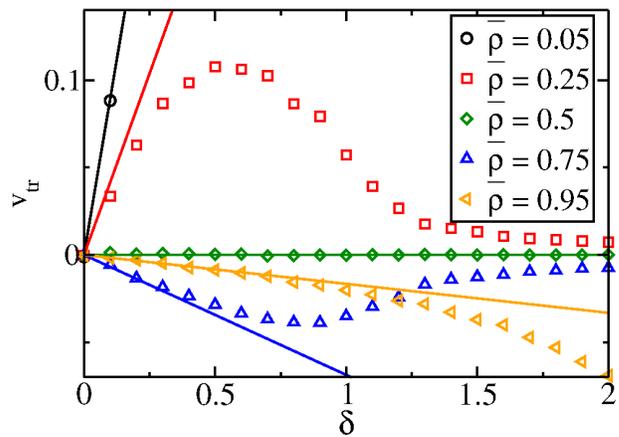} 
	\end{center}
	\caption{\small Velocity of the tracer as a function of $\delta$ for fixed $r=1$, $\rp=0.5$ and $\deltap=0$ and various densities in a system of length $L=500$. For large densities the sign of the velocity is opposite to the one of $\delta$. Note also the NDM occurring for $\rhob = 0.25$ and $\delta \gtrsim 0.5$. Numerical results (symbols) are compared to the theory of Section~\ref{section:linres} for small $\delta$ (lines).}  
	\label{fig:vtrd}
\end{figure}

\begin{figure} 
	\begin{center}
		\includegraphics[width=0.45\textwidth]{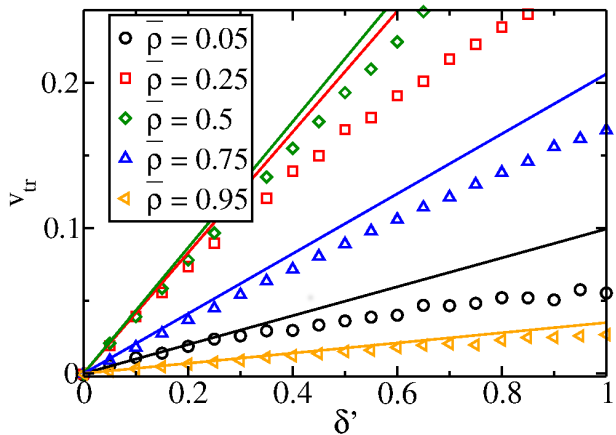} 
	\end{center}
	\caption{\small Velocity of the tracer as a function of $\deltap$ for fixed $r=1$, $\rp=0.5$ and $\delta=0$ and various densities in a system of length $L=500$. 
		Numerical results (symbols) are compared to the theory of Section~\ref{section:linres} for small $\deltap$ (lines).}  
	\label{fig:vtrdp}
\end{figure}

\subsection{Numerical results}

The system defined above can be easily simulated for any values of $p = r + \frac{\delta}{2}$, $q = r -\frac{\delta}{2}$, $\pp = \rp + \frac{\deltap}{2}$ and $\qp = \rp - \frac{\deltap}{2}$. In Figures~\ref{fig:vtrd} and~\ref{fig:vtrdp} we present numerical results of the tracer velocity for fixed values of $r$ and $r'$ as a function of the respective biases, first $\delta \neq 0$ and then $\deltap \neq 0$. In particular, for $\delta \neq 0$ and $\deltap = 0$ (Fig.~\ref{fig:vtrd}), the curves are monotonously increasing for small densities, but start to exhibit NDM and even ANM for larger densities. On the contrary, for $\delta = 0$ and $\deltap \neq 0$ the curves are monotonously increasing (Fig.~\ref{fig:vtrdp}).

\begin{figure} 
	\begin{center}
		\includegraphics[width=0.45\textwidth]{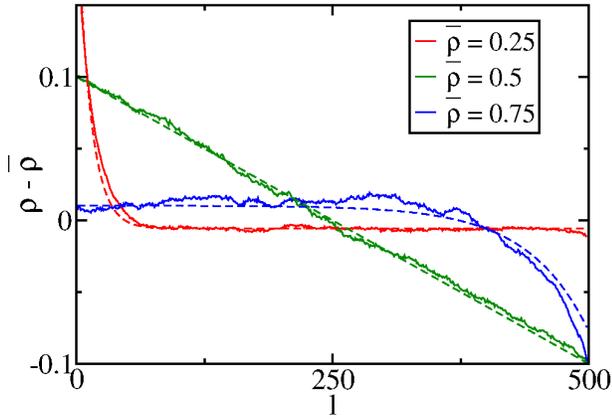} 
	\end{center}
	\caption{\small Density profiles in the tracer frame for different average densities $\rhob$ in systems with $r=1$, $\rp=0.5$, $\delta=0.4$ and $\deltap = 0$. Numerical results (solid lines) are compared to the
		predictions of section~\ref{section:density} (dashed lines). The decay length~\eqref{eq:decaylngth} changes sign at $\rho = 0.5$, which goes hand in hand with the change in the sign of the velocity observed in Figure~\ref{fig:vtrd}.}  
	\label{fig:dend}
\end{figure}

\begin{figure} 
	\begin{center}
		\includegraphics[width=0.45\textwidth]{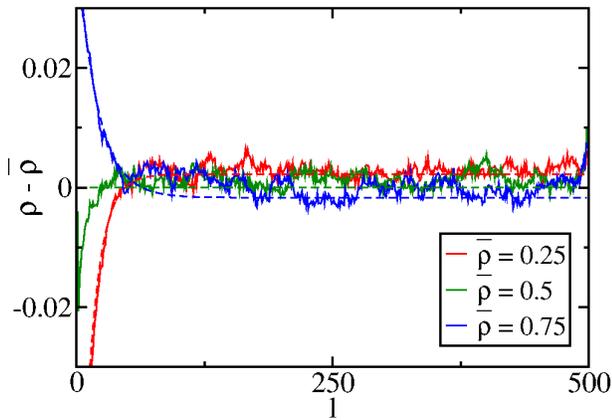} 
	\end{center}
	\caption{\small Density profiles in the tracer frame for different average densities $\rhob$ in systems with $r=1$, $\rp=0.5$, $\delta=0$ and $\deltap = 0.4$. Numerical results (solid lines) are compared to the
		predictions of section~\ref{section:density} (dashed lines). The density difference between the right and the left of the tracer changes sign at $\rhob = 0.5$, see the theoretical expression~\eqref{eq:solphi} with the simplification~\eqref{eq:signC1} for the prefactor.
	}  
	\label{fig:dendp}
\end{figure}

In Figures~\ref{fig:dend} and~\ref{fig:dendp} we plot the corresponding density profiles for different values of the average density. The density is found to be flat in the bulk of the system, \textit{i.e.} far from the tracer, with a meniscus appearing on one side of the tracer. It appears that the change of sign in the velocity for $\delta \neq 0$ as the density is increased is accompanied by a qualitative change in the density profile, where a meniscus appears at the front of the tracer for low densities and at its back for high densities (see Fig.~\ref{fig:dend}). 


In the next subsection we study the system around its equilibrium state using a mean-field approximation, and compute $\vtr$ in the linear response regime.

\subsection{Tracer velocity}
\label{section:linres}

We start by writing mean-field equations for the densities $\{\rho_l\}_{l=1,\ldots,L-1}$,
\begin{widetext}
\begin{eqnarray}
\label{eq:mfdensity}
\frac{\dd \rho_1}{\dd t} &=& \rho_2 -\rho_1 + p (1-\rho_1) \rho_2 - q (1-\rho_{L-1} ) \rho_1 +\pp (1-\rho_1) \rho_2 \rho_3 - \qp (1-\rho_{L-1}) \rho_{L-2} \rho_1, \nonumber \\
\frac{\dd \rho_2}{\dd t} &=& \rho_3 -2 \rho_2 + \rho_1 + p (1-\rho_1) (\rho_3 - \rho_2) + q (1-\rho_{L-1} ) (\rho_1-\rho_2) \nonumber \\ && +\pp (1-\rho_1) \rho_2 (\rho_4 -1) + \qp (1-\rho_{L-1}) \rho_{L-2} (1-\rho_2), \nonumber \\
\frac{\dd \rho_l}{\dd t} &=& \rho_{l+1} -2 \rho_l + \rho_{l-1} + p (1-\rho_1) (\rho_{l+1} - \rho_l) + q (1-\rho_{L-1} ) (\rho_{l-1}-\rho_l) \\ && +\pp (1-\rho_1) \rho_2 (\rho_{l+2} -\rho_l) + \qp (1-\rho_{L-1}) \rho_{L-2} (\rho_{l-2}-\rho_l), \qquad \qquad l=3,\ldots, L-3, \nonumber \\
\frac{\dd \rho_{L-2}}{\dd t} &=& \rho_{L-1} -2 \rho_{L-2} + \rho_{L-3} + p (1-\rho_1) (\rho_{L-1} - \rho_{L-2}) + q (1-\rho_{L-1} ) (\rho_{L-3}-\rho_{L-2}), \nonumber \\ && + \pp (1-\rho_1) \rho_2 (1 - \rho_{L-2}) + \qp (1-\rho_{L-1}) \rho_{L-2} (\rho_{L-4}-1), \nonumber \\
\frac{\dd \rho_{L-1}}{\dd t} &=& -\rho_{L-1} + \rho_{L-2} - p (1-\rho_1) \rho_{L-1} + q (1-\rho_{L-1} ) \rho_{L-2} - \pp (1-\rho_1) \rho_2 \rho_{L-1} + \qp (1-\rho_{L-1}) \rho_{L-2} \rho_{L-3}, \nonumber
\end{eqnarray} 
where we factorized correlations $\langle \tau_{l_1} \ldots \tau_{l_k}\rangle = \rho_{l_1} \ldots \rho_{l_k}$ for distinct positions. Since correlations are factorized in equilibrium, it is reasonable to expect that this approximation will give good results at least close to equilibrium. A similar technique has also been proven accurate in closely related systems, see \textit{e.g.}~\cite{benichou1999, deconinck_o_m1997, benichou2001, benichou2000b, benichou2015, brummelhuis_h1989, illien2013, illien2015, benichou2013a,benichou2013b,benichou2013c, benichou2016}. The tracer velocity~\eqref{eq:vtrdef} becomes
\begin{eqnarray}
\label{eq:vtrmf}
\vtr &=&  p (1-\rho_1) -q (1- \rho_{L-1}) + 2 \pp (1-\rho_1) \rho_2 - 2 \qp (1-\rho_{L-1}) \rho_{L-2}.
\end{eqnarray}
\end{widetext}

We consider equations~\eqref{eq:mfdensity} in the stationary state $\dd \rho_l/\dd t = 0$. Because of particle number conservation, they give only $L-2$ independent conditions. The missing $(L-1)^\mathrm{th}$ condition is obtained by fixing the number of particles,
\begin{equation}
\label{eq:norm}
\sum_{l=1}^{L-1} \rho_l = N.
\end{equation}
When there is no bias, $\delta = \delta' = 0$, it is easy to confirm that a flat density profile $\rho_l = \rhob$ solves equations~\eqref{eq:mfdensity}-\eqref{eq:norm} so that the tracer velocity~\eqref{eq:vtrmf} vanishes. 
 
 For simplicity, we now consider the case where the biases $\delta$ and $\delta'$ are small and of the same order. We expand the density,
\begin{equation}
\label{eq:defsigma}
\rho_l = \rhob + \delta \sigma_l + \delta' \sigma_l' + \cO(\delta^2),
\end{equation}
and study the solution to linear order in $\delta$ and $\deltap$. While the tracer velocity is rather well-described to this order, this is not the case for the density profile. In order to obtain the density profile, one has to study the equations to second order in $\delta$ and $\deltap$. This will be done in the next subsection.

We start by solving the equations for the bulk sites $l=3,\ldots,L-3$. The terms linear in $\delta$ and $\delta'$ give equations for $\sigma_l$ and $\sigma_l'$, respectively. Both bulk equations turn out to be the same,
\begin{eqnarray}
\label{eq:bulk}
(1+r (1-\rhob)) (\sigma_{l+1}-2 \sigma_l+\sigma_{l-1}) && \\ + r' \rhob (1-\rhob) (\sigma_{l+2}-2 \sigma_l+\sigma_{l-2}) &=& 0,\nonumber 
\end{eqnarray}
for $l=3,\ldots,L-3$, and the very same equation for the $\{\sigma_l'\}_{l=1,\dots,L-1}$. Note that in the continuum limit this equation would reduce to a Laplace equation $\partial_l^2 \sigma = 0$, giving a linear density profile.
The solution of the discrete equation is
\begin{eqnarray}
\label{eq:solbulk}
\sigma_l &=& \alpha + \beta l +\gamma_+ X^l + \gamma_- X^{L-l}, \nonumber \\
\sigma_l' &=& \alpha' + \beta' l +\gamma_+' X^l + \gamma_-' X^{L-l},
\end{eqnarray}
for $l=1,\ldots,L-1$, where 
\begin{eqnarray}
\label{eq:Xdef}
X &=& - \left( 1+ \frac{1+r (1-\rhob)}{2 \rp \rhob (1-\rhob)}\right) \\ &&+ \sqrt{\left( 1+ \frac{1+r (1-\rhob)}{2 \rp \rhob (1-\rhob)}\right)^2-1} \nonumber 
\end{eqnarray}
and $X^{-1}$ are the roots of 
\begin{eqnarray}
\label{eq:Xeq}
(1+r(1-\rhob)) X +\rp \rhob (1-\rhob) (1+X)^2 = 0.
\end{eqnarray} 
Note that $|X| < 1$.

We now have to satisfy the boundary and normalization conditions. In the large $N$ and $L$ limit with $\frac{N}{L-1} = \rhob$, the normalization~\eqref{eq:norm}  gives $\beta = - \frac{2 \alpha}{L}$ and $\beta' = - \frac{2 \alpha'}{L}$. Let us now take the sum of the equations for $\rho_1$ and $\rho_{L-1}$. Sorting the $\delta$ and $\delta'$ terms, we again get the same equation for $\sigma_l$ and $\sigma_l'$,
\begin{eqnarray}
\label{eq:s1Lm1}
 (1+r(1-\rhob)) (\sigma_2 - \sigma_1 + \sigma_{L-2} - \sigma_{L-1})&& \\ + \rp \rhob (1-\rhob) (\sigma_3 - \sigma_1 + \sigma_{L-3} - \sigma_{L-1}) &=& 0. \nonumber
\end{eqnarray}
Taking the sum of the equations for $\rho_2$ and $\rho_{L-2}$ 
leads to the same result.
For large $L$, we have that $\sigma_l \simeq \alpha + \gamma_+ X^l$ for $l=\cO(1)$, and $\sigma_{L-l} \simeq -\alpha + \gamma_- X^l$ for $L-l=\cO(1)$. Inserting these forms in equation~\eqref{eq:s1Lm1}, we get $\gamma_- = - \gamma_+$ and, similarly, $\gamma_-' = - \gamma_+'$. The linear perturbations to the densities therefore have the form
\begin{eqnarray}
\label{eq:solbulks}
\sigma_l &=& \alpha \left(1- 2 \frac{l}{L} \right) +\gamma_+ \left( X^l - X^{L-l} \right), \nonumber \\
\sigma_l' &=& \alpha' \left(1- 2 \frac{l}{L} \right) +\gamma_+' \left( X^l - X^{L-l} \right).
\end{eqnarray}
The equations for, say, $\rho_1$ and $\rho_2$ give two systems of two linear equations for $\alpha$ and $\gamma_+$, and for $\alpha'$ and $\gamma_+'$. The solutions to these equations are obtained using Mathematica and are given in Appendix~\ref{section:appagag}. 

The density profile obtained in this analysis is linear in the bulk with exponential layers on both sides of the tracer. This is different from the numerical observation of a flat profile in the bulk and an exponential layer only on one side of the tracer. This discrepancy is a result of the fact that the analysis has been carried out to linear order in $\delta$ and $\deltap$. This will be corrected in Section~\ref{section:density}.

The velocity of the tracer can be obtained from the mean-field expression~\eqref{eq:vtrmf} and the solutions~\eqref{eq:solbulks}, where the coefficients are given by~\eqref{eq:ag}. We separate it into two contributions, namely the one coming from the hops of the tracer towards an empty site (terms proportional to $p$ and $q$ in
equation~\eqref{eq:vtrmf}) and the one coming from exchanges (terms proportional to $\pp$ and $\qp$ in  
the same equation). We can write $\vtr = \vtrh + \vtre$, where the subscripts $\text{H}$ and $\text{E}$ indicate the contributions coming from hops and exchanges, respectively. Each of these pieces has a term proportional to its 
corresponding bias,
\begin{equation}
\vtr_{,A} = \mu_{A,\text{H}} \delta + \mu_{A,\text{E}} \deltap,
\end{equation}
 which gives four coefficients $\mu_{A,B}$ with $A,B \in \{\text{H},\text{E}\}$. Explicitly, they are
\begin{widetext}
 \begin{eqnarray}
 \label{eq:musol}
 \mu_{\text{H},\text{H}} &=& (1-\rhob) - r (\sigma_1 - \sigma_{L-1}) = (1-\rhob) - 2 r (\alpha + \gamma_+ X) = \frac{\rp (1-2 \rhob)^2}{2 r \rhob^2} \muee, \nonumber \\
  \mu_{\text{H},\text{E}} &=& - r (\sigma_1' - \sigma_{L-1}') = - 2 r (\alpha' + \gamma_+' X) = \frac{1-2 \rhob}{2 \rhob} \muee, \\
  \mu_{\text{E},\text{H}} &=& 2 \rp \left( (1-\rhob) (\sigma_2 - \sigma_{L-2}) + \rhob (\sigma_{L-1} - \sigma_1)\right) = 4 \rp \left( (1-\rhob) (\alpha + \gamma_+ X^2) - \rhob (\alpha + \gamma_+ X) \right) = \frac{\rp (1-2 \rhob)}{ r \rhob} \muee, \nonumber
  \end{eqnarray}
  with the last coefficient
  \begin{eqnarray}
 \label{eq:muee}
  \muee &=& 2 \rhob (1-\rhob) + 2 \rp (1-\rhob) (\sigma_2' - \sigma_{L-2}') + 2 \rp \rhob (\sigma_{L-1}'-\sigma_1') = 2 \rhob (1-\rhob) + 4 \rp (1-\rhob) (\alpha' + \gamma_+' X^2) - 4 \rp \rhob (\alpha' + \gamma_+' X) \nonumber \\
 &=& \frac{2 \rhob^2 (1-\rhob) r \sqrt{1+r (1-\rhob)}}{(2 \rhob - 1) (r+\rp (2 \rhob - 1)) \sqrt{1+r (1-\rhob)} + r (1-\rhob) \sqrt{1+r(1-\rhob) + 4 \rp \rhob (1-\rhob)}}.
 \end{eqnarray}
\end{widetext}
In Appendix~\ref{section:appmuee} we show that $\muee > 0$ for all $0 < \rhob < 1$, $r > 0$, $\rp > 0$. 

For clarity, let us group the linear response coefficients~\eqref{eq:musol}-\eqref{eq:muee} into a linear response matrix,
\begin{equation}
\label{eq:linresmat}
\left( \begin{array}{c} \vtrh \\ \vtre \end{array} \right) 
= \frac{\rp \muee}{2 \rhob^2} \left( \begin{array}{cc} (1-2\rhob)^2 & 2 \rhob (1-2\rhob) \\ 2 \rhob (1-2\rhob) & 4 \rhob^2 \end{array} \right) \left( \begin{array}{c} \delta/r \\ \deltap/(2\rp) \end{array} \right),
\end{equation}
where the entries of the column vector on the RHS are the thermodynamically conjugate forces. In this basis the response matrix is symmetric, as expected from the Onsager relations.
In~\eqref{eq:linresmat} we see that the diagonal coefficients of the response matrix are always positive, consistent with fluctuation-dissipation relations. Conversely, the off-diagonal coefficients need not be positive and indeed they change sign at $\rhob = \frac{1}{2}$, which allows for ANM. Thus ANM found in this model in the linear response regime is a direct result of the fact that the dynamics involves two driving mechanisms, namely hopping ($p/q$) and exchange ($\pp/\qp$). 
Note that the columns of the linear response matrix are proportional, which shows that the response of the tracer to the two driving fields is the same. This indicates that exchange and hopping are completely coupled in the sense of~\cite{kedem_c1965}. 
The total velocity becomes
\begin{equation}
\label{eq:linresvtr}
\vtr = \frac{\muee}{2 r \rhob^2} \left( \rp (1-2\rhob) \delta + r \rhob \deltap \right).
\end{equation}
For fixed $\delta, \deltap > 0$, the velocity starts out positive for small $\rhob$, and changes sign for $\rhob = \left(2 - \frac{r \deltap}{\rp \delta} \right)^{-1}$, which is smaller than $1$ for $r \deltap < \rp \delta$. 
The prediction~\eqref{eq:linresvtr} is compared to the results of Monte-Carlo simulations in Figures~\ref{fig:vtrd} and~\ref{fig:vtrdp}, and the agreement is very good.

\begin{figure}[!ht]
	\begin{center}
		\includegraphics[width=0.45\textwidth]{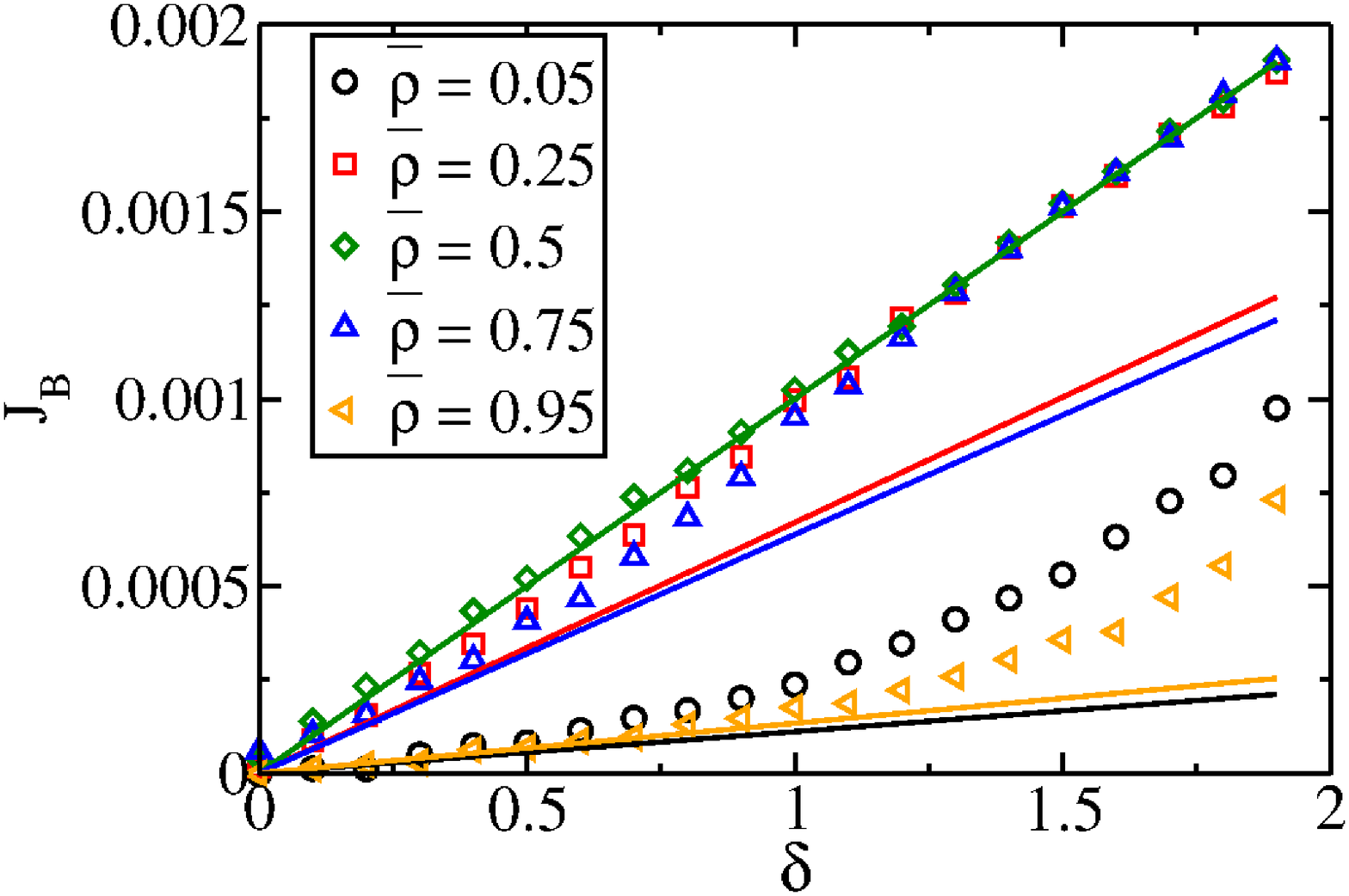} 
	\end{center}
	\caption{\small The bath particle current - for the same parameters as in Fig.~\ref{fig:vtrd} - are compared 
	to the analytical predictions for small $\delta$~\eqref{eq:Jb}.}  
	\label{fig:Jbd}
\end{figure}

\begin{figure}[!ht]
	\begin{center}
		\includegraphics[width=0.45\textwidth]{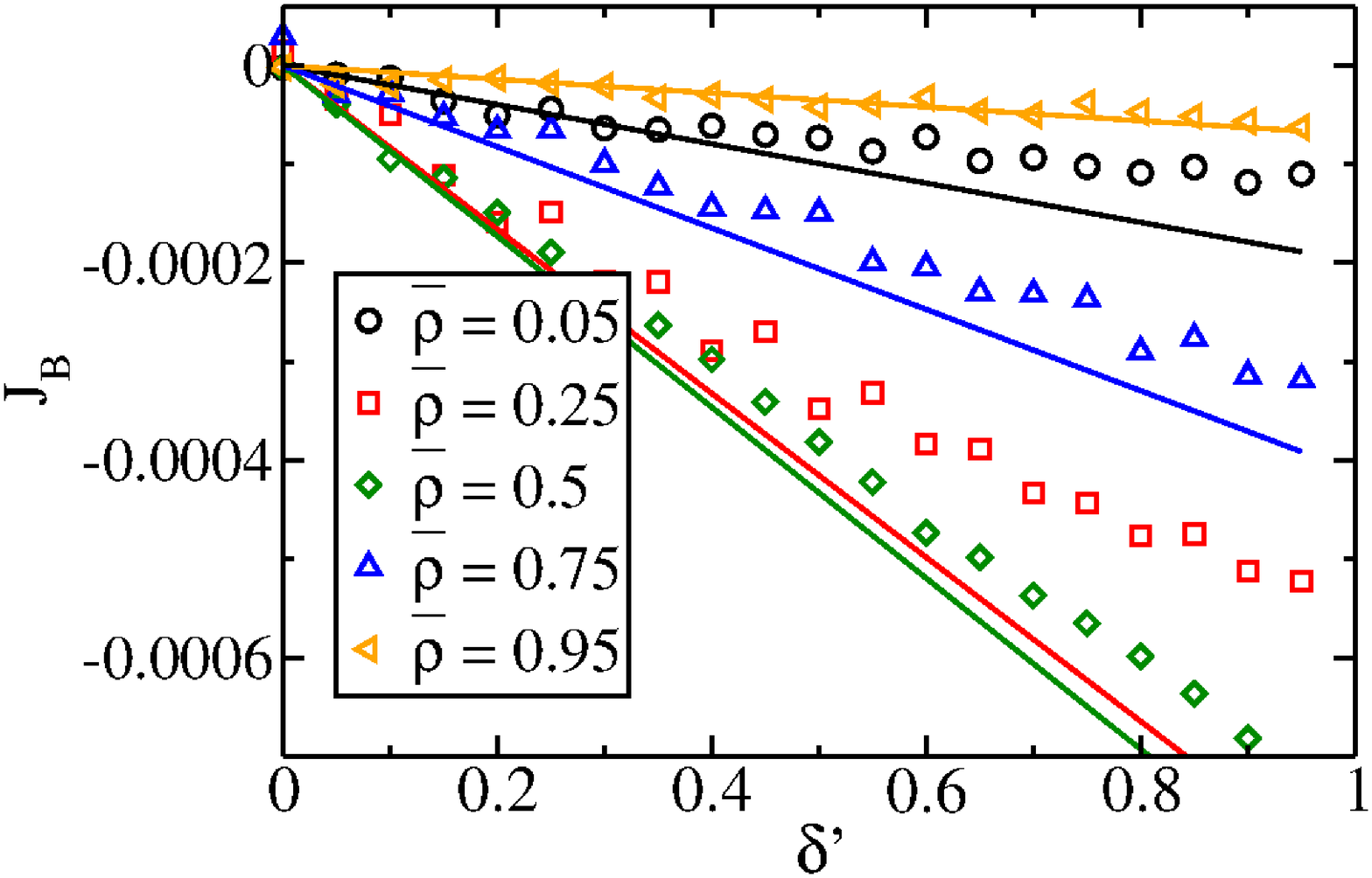} 
	\end{center}
	\caption{\small The bath particle current  - for the same parameters as in Fig.~\ref{fig:vtrdp} - are compared 
	to the analytical prediction for small $\deltap$~\eqref{eq:Jb}.}  
	\label{fig:Jbdp}
\end{figure}

Similarly, one can predict the current $\Jb$ of bath particles in the linear regime. It can be expressed as a function of the densities in the neighborhood of the tracer (see Appendix~\ref{section:appJb}), 
\begin{equation}
\label{eq:Jb}
\Jb = \frac{\rho_1 - \rho_{L-1} + 2 \qp (1-\rho_{L-1}) \rho_{L-2} - 2 \pp (1-\rho_1) \rho_2}{L}.
\end{equation}
Using the computed values of $\rho_1$, $\rho_2$, $\rho_{L-2}$ and $\rho_{L-1}$,,  one obtains analytical predictions for $\Jb$. They are compared to Monte Carlo simulations in Figures~\ref{fig:Jbd} and~\ref{fig:Jbdp}. The slope at the origin is predicted accurately.

Note that the analysis presented in this section yields good agreement for $\vtr$ and $\Jb$, since they are determined by the average density of the sites close to the tracer (Eqs.\,\eqref{eq:vtrmf},\eqref{eq:Jb}). The total current can be obtained as $\frac{\vtr}{L} + \Jb$ and is therefore predicted accurately by the linear analysis as well. 

While these densities are well described by Eqs.\,\eqref{eq:solbulk},\eqref{eq:Xdef}, the overall density profile is not.
Let us now focus on correcting the discrepancy in the density profile by extending the analysis to second order in $\delta$ and $\deltap$.

\subsection{Density profile}
\label{section:density}

In order to explain the form of the density profile for small biases, one has to keep higher-order terms in the expansion. We consider a system with small biases $\delta$ and $\deltap$ and write $\phi_l = \rho_l - \rhob$. Expanding the bulk equation to second order in $\phi_l$ gives
\begin{widetext}
\begin{eqnarray}
\label{eq:phi}
0 &=& (1+r(1-\rhob)) (\phi_{l+1} - 2 \phi_l + \phi_{l-1}) + \rp \rhob (1-\rhob) (\phi_{l+2} - 2 \phi_l + \phi_{l-2}) \nonumber \\ &&+ \left( (1-\rhob) \frac{\delta}{2} - r \phi_1 \right) (\phi_{l+1} - \phi_l) + \left( -(1-\rhob) \frac{\delta}{2} - r \phi_{L-1} \right) (\phi_{l-1} - \phi_l) \\ && + \left( (1-\rhob) \rhob \frac{\deltap}{2} - \rp \rhob \phi_{1} + \rp (1-\rhob) \phi_{2} \right) (\phi_{l+2} - \phi_l) + \left( -(1-\rhob) \rhob \frac{\deltap}{2} - \rp \rhob \phi_{L-1} + \rp (1-\rhob) \phi_{L-2} \right) (\phi_{l-2} - \phi_l) \nonumber \\
&\simeq& \left[1+r(1-\rhob) + 4 \rp \rhob (1-\rhob)\right] \partial_l^2 \phi + \left[ (1-\rhob) \delta + 2 \rhob (1-\rhob) \deltap - (r+2\rp \rhob) (\phi_1 - \phi_{L-1}) + 2 \rp (1-\rhob) (\phi_2 - \phi_{L-2}) \right] \partial_l \phi, \nonumber 
\end{eqnarray}
\end{widetext}
where we approximated discrete differences by derivatives. The coefficient of the first order derivative is exactly the part of $\vtr$ linear in $\phi$, and it depends on values of $\phi$ close to the tracer.
Let us replace them by their values from the preceding section, 
\begin{equation}
\label{eq:approxphi}
\phi_n - \phi_{L-n} \simeq 2 \delta (\alpha + \gamma_+ X^n) + 2 \deltap (\alpha' + \gamma_+' X^n),
\end{equation}
for $n=1,2$, so that the aforementioned coefficient exactly becomes the $\vtr$ as obtained in equation~\eqref{eq:linresvtr}. The solution of~\eqref{eq:phi} is exponential, 
\begin{equation}
\label{eq:solphi0}
\phi_l = C_1 \ee^{-\frac{l}{\xi}} + C_2,
\end{equation}
where $C_1$ and $C_2$ are integration constants, and the decay length is
\begin{equation}
\label{eq:decaylngth}
\xi = \frac{1+r(1-\rhob) + 4 \rp \rhob (1-\rhob)}{\vtr}.
\end{equation} 
We use the definition~\eqref{eq:decaylngth}, where $\xi$ can be either positive or negative, to make the presentation simpler.

The decay length is of order $\delta^{-1}$ or $\deltap^{-1}$, which explains why we found linear profiles when we neglected $\cO(\delta^2)$ terms. One can check that expression~\eqref{eq:decaylngth} diverges at $\rhob=\frac{1}{2}$ for $\deltap = 0$, and at $\rhob=1$ for any $\delta$,$\deltap$. This is consistent with the respective linear profiles obtained in Fig.\,\ref{fig:dend} and the high-density calculation of Section~\ref{section:highrho}. In the low-density limit $\rhob \to 0$ one simply gets $\xi \sim \frac{1+r}{\delta}$, which is the diffusion coefficient of the free tracer divided by the bias.

The constants $C_1$ and $C_2$ are linked by mass conservation, $\sum_{l=1}^{L-1} \phi_l = 0$, giving 
\begin{equation}
\label{eq:consmC1C2}
C_2 = - \frac{1}{L-1} \frac{\ee^{-\xi^{-1}} - \ee^{-\xi^{-1} L}}{1-\ee^{-\xi^{-1}}} C_1.
\end{equation}
 As $C_1 = \left( \ee^{-\xi^{-1}} - \ee^{-\xi^{-1}(L-1)} \right)^{-1} \left( \phi_1 - \phi_{L-1}\right)$, we may employ approximation~\eqref{eq:approxphi} again in order to obtain the value of $C_1$. We end up with 
\begin{eqnarray}
\label{eq:solphi}
\phi_l &=& \frac{2 \left[ (\alpha+\gamma_+ X) \delta + (\alpha'+\gamma_+' X) \deltap \right]}{1 - \ee^{-\xi^{-1}(L-2)}} \\ && \times \left[ \ee^{-\xi^{-1} (l-1)} - \frac{1}{L-1} \frac{1 - \ee^{-\xi^{-1}(L-1)}}{1-\ee^{-\xi^{-1}}} \right]. \nonumber 
\end{eqnarray}
The form~\eqref{eq:solphi} is shown to be in good agreement with simulations in Figs.\,\ref{fig:dend} and~\ref{fig:dendp}.

\subsection{High density regime}
\label{section:highrho}

Here we go beyond linear response for high densities $\rhob \simeq 1$. In that case the holes are very sparse and can be considered independent, so that we can start by studying a system with $L-2$ bath particles and only one hole. Let $l = 1,\ldots,L-1$ denote the position of the hole with respect to the tracer. Examination shows that the probability distribution of the position of the hole, $P_l(t)$, obeys
\begin{eqnarray}
\label{eq:meqhole}
\frac{\dd P_l}{\dd t} &=& P_{l+1} - 2 P_l + P_{l-1} \\ && + \delta_{l,1} [(1-p-\pp)P_1 + (q+\qp) P_{L-1}] \nonumber \\ && + \delta_{l,L-1} [ (1-q-\qp)P_{L-1} + (p+\pp) P_1 ], \nonumber
\end{eqnarray}
with the convention $P_0 = P_L = 0$, and the normalization $\sum_{l=1}^{L-1} P_l = 1$. When the hole is far from the tracer, $l \neq 1, L-1$, it simply diffuses as seen on the first line of~\eqref{eq:meqhole}. The two other terms in this equation correspond to hopping and exchange processes which take place when the hole is next to the tracer. 

In the stationary state and large $L$ limit, these equations give
\begin{equation}
\label{eq:solPY}
P_l = \frac{2}{L(p+\pp+q+\qp)} \left( (p+\pp-q-\qp) \frac{l}{L} +q+\qp \right).
\end{equation}
This gives, for one hole,
\begin{equation}
\label{eq:linresmathighole}
\left( \begin{array}{c} \vtrh^{(1)} \\ \vtre^{(1)} \end{array} \right) 
= \left( \begin{array}{c} p P_1 - q P_{L-1} \\ 2 \pp P_1 - 2 \qp P_{L-1} \end{array} \right) 
= \frac{r \deltap - \rp \delta}{L (r + \rp)} \left( \begin{array}{c} -1 \\ 2 \end{array} \right),
\end{equation}
to all orders in $\delta$ and $\deltap$. For a system with not too large a number of holes $L (1-\rhob)$, we can simply add the effect of each hole. This gives
\begin{equation}
\label{eq:vtrhighrho}
\vtr = - \vtrh = \frac{\vtre}{2} = (1-\rhob) \frac{r \deltap - \rp \delta}{r + \rp},
\end{equation}
and the density profile is linear,
\begin{equation}
\label{eq:denhighrho}
\rho_l = \rhob-  (1-\rhob) \frac{\delta+\deltap}{r +\rp} \left(\frac{l}{L} - \frac{1}{2}\right).
\end{equation}
The results~\eqref{eq:vtrhighrho} and~\eqref{eq:denhighrho} are expected to be exact in the high-density limit.
They also match the high-density limits of the mean-field predictions for the velocity~\eqref{eq:linresvtr} and the density profile~\eqref{eq:solphi}.
Indeed, the agreement between the predicted profile~\eqref{eq:denhighrho} and 
the numerical results can be checked to be very good for large densities.

\begin{figure}[!ht]
	\begin{center}
		\includegraphics[width=0.48\textwidth]{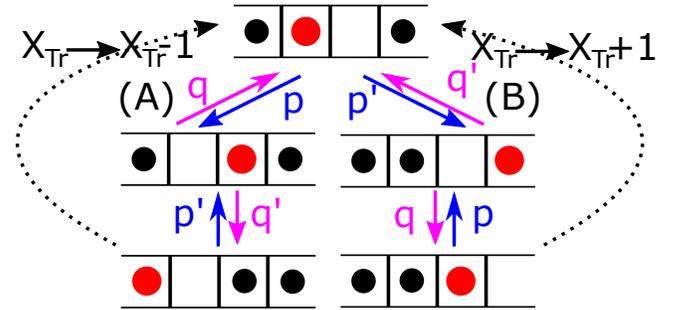} 
	\end{center}
	\caption{\small Mechanism leading to ANM in a system with one hole. Far from the tracer, the hole diffuses symmetrically and will at some point reach a site neighboring the tracer, say to its right. In that case, the tracer may either hop forward (A) or exchange positions with the next bath particle (B). Option (B) eventually brings the tracer back to its starting state shifted one site to the right and is therefore the strategy to make progress to the right. On the other hand, after another exchange step option (A) also brings the tracer back to its starting position, but this time shifted one step to the left. For $p \qp > q \pp$ the preferred route is (A), which leads to $\vtr < 0$.}  
	\label{fig:mechanm}
\end{figure}

More importantly, considering a system with one hole helps to shed light on the way ANM occurs, see Fig.\,\ref{fig:mechanm} and the explanation in the caption. 
The sequence of transitions shown in Fig.\,\ref{fig:mechanm} contains a step where the tracer hops to the right and is therefore favored by an increase of $p$. The net result of this sequence is an overall displacement of the system to the left. Symmetrically, an increase of $q$ favors a sequence of transitions that 
results in a net displacement of the tracer to the right. Therefore, when $p > q$ the tracer moves preferentially to the left and ANM is observed.

\section{Brownian hard disks in a narrow channel}
\label{section:hd}


Motivated by the lattice model of driven tracer presented in the preceding section, we move a small step in the direction of a more realistic system and consider
a gas of hard disks in a planar narrow channel. Even though the fluctuation-dissipation relation forbids ANM in linear response, the gas could in principle exhibit ANM at large driving field. The channel is oriented parallel to the
$x$ axis with the origin of the coordinate system in its center. It is periodic in 
$x$ with periodicity $L_x$. The channel width $L_y' = 2 \sigma + \epsilon $ is
chosen to allow passing and overtaking of the particles with diameter $\sigma$. Thus, $\epsilon > 0$
is assumed.
In this channel there are $N-1$ neutral particles with mass $m_k$, position $\mathbf{r}_k = (x_k,y_k)$,
and velocity $\mathbf {v}_k = (v_{k,x},v_{k,y})$,  $k \in \{1,2,\cdots,N-1\}$. In addition, there is a
tracer particle with index $k = 0$, which is  driven by a homogeneous force  ${\bf F} = (F,0)$ 
parallel to the channel axis.
The dynamics of all disks is assumed to follow an underdamped Langevin equation,
\begin{eqnarray}
	\frac{ d \mathbf{r}_k}{dt} & =  &\mathbf{v}_k \label{eom1} \\
	m_k \frac{ d \mathbf{v}_k}{dt} & = & 
	\mathbf{F}  \delta_{k,0}  -\gamma \mathbf{v}_k +   \sqrt{ 2 \gamma k_B T} \xi_0, \label{eom2} + \{\mbox{coll.}\}\\
	\nonumber
\end{eqnarray}
for $ k \in \{0,1,\cdots,N-1 \}$. As usual,  $\xi$ is a delta-correlated Gaussian white noise,  
$\delta_{k,0} $ is unity for $k = 0$ and zero otherwise, $k_B$ is the
Boltzmann constant, and $T$ is the temperature of the bath.

The stochastic motion equations are solved to first order in the time step $\Delta t$ according to the updating formulas of Gillespie~\cite{gillespie1996a, gillespie1996b}. Reduced units are used
throughout, for which the mass of the tracer, $m_0$, the particle diameter $\sigma$ and the 
energy $k_B T$ are unity. In these units, the Langevin friction parameter $\gamma$, which
determines the noise strength, is set to $2$ and the time step $\Delta t$ to $10^{-3}$. In all simulations below, the following parameters are 
used: channel length $L_x = 300$, total number of particles $N = 200$, and tracer mass $m_0 = 1.0$. The 
masses $m_k \equiv m$ of the neutral particles $k = 1,\cdots, 199$ and the driving force $F$ are indicated 
where needed. 

There are two kinds of collisions, namely  collisions between particles and collisions of a particle with a
long hard boundary of the channel. Particle-particle collisions are strictly elastic. The long channel boundaries, however, 
are thermal van Beijeren walls~\cite{vanbeijeren2014}, which re-emit colliding particles in equilibrium with the
boundary temperature $T_b$. The latter is taken to agree with the bath temperature, $T_b = T$.
Since in such a narrow channel the long thermal walls are of significant importance for the 
non-equilibrium transport, a short description of the van Beijeren walls used here is given in
Appendix~\ref{section:appvbthermostat}.

\begin{figure}[!ht]
	\begin{center}
		\includegraphics[width=0.5\textwidth]{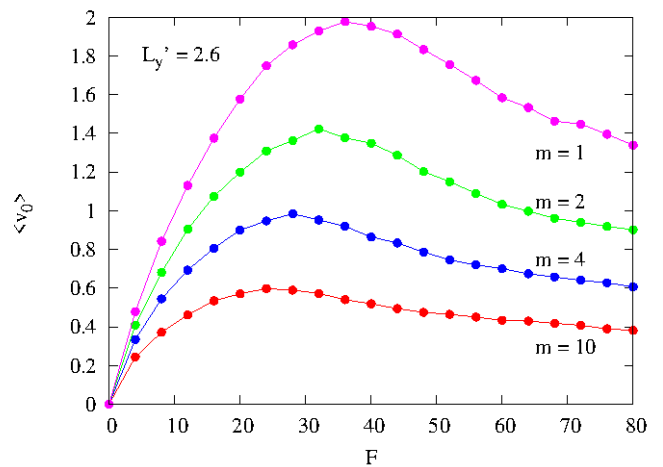} 
	\end{center}
	\caption{\small   Dependence of the mean tracer velocity on the homogeneous  
		force  $F$ for channels of  width $L_y' = 2.6$. 
		 The  masses  of the neutral 
		particles differ for the curves as indicated by the labels.}
	\label{NDM}
\end{figure}

The dependence of the  mean tracer speed $<v_0>$ on the drive $F$ is shown in Fig. \ref{NDM}
for a channel of width $L_y' = 2.6$. The
four curves  are for different masses of the neutral particles as indicated by the labels. For low $F$ the velocity increases  almost linearly with $F$ (Ohm's law). For
larger fields, however, the curves bend over and reach a regime with a negative slope indicating
negative differential mobility.
This behavior  is most prominent for $2.3 <  L_y' < 2.8 $  and deteriorates quickly for broader channels.
This is demonstrated by a comparison  of the
\begin{figure}[!ht]
	\begin{center}
		\includegraphics[width=0.5\textwidth]{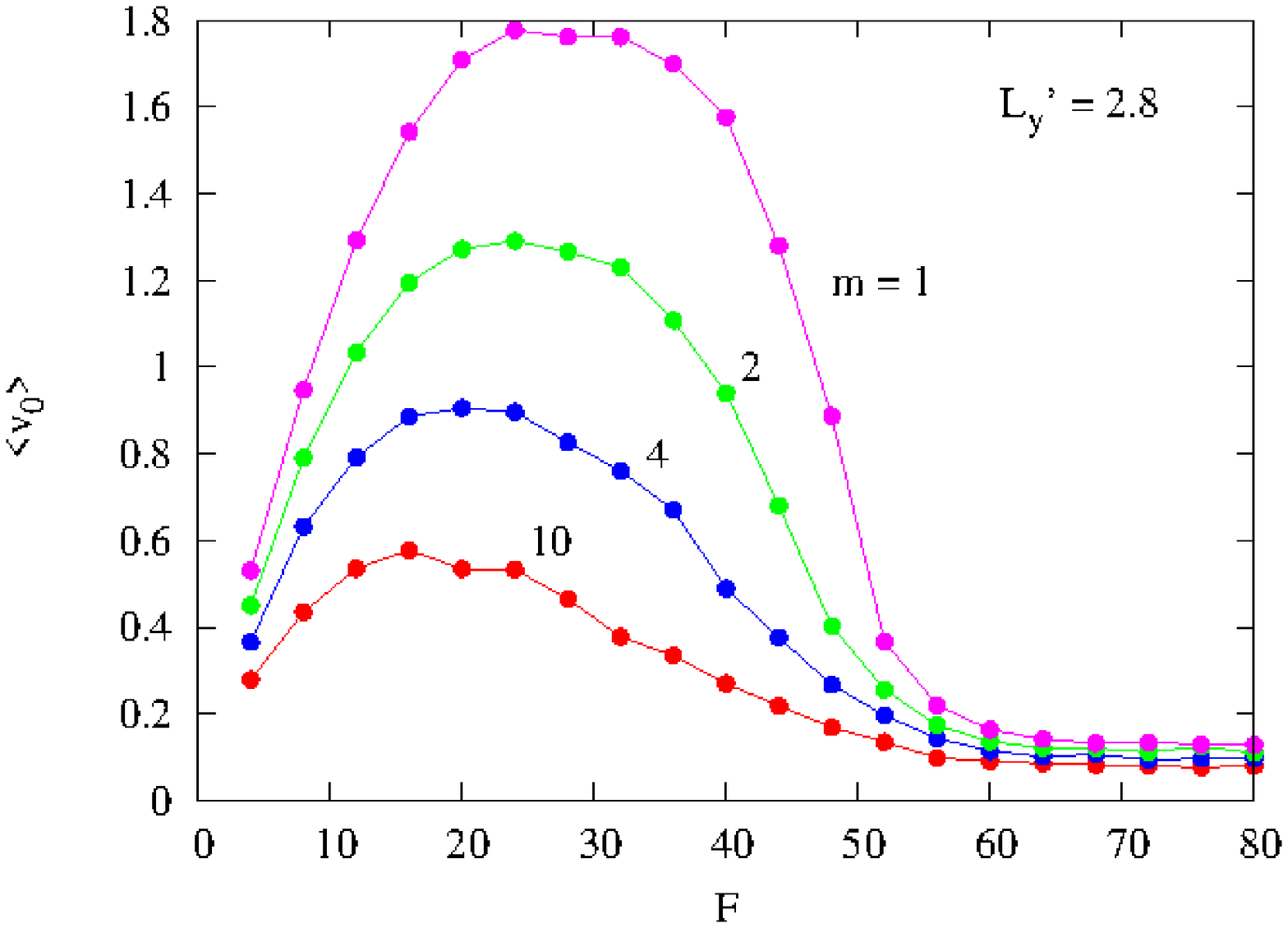} \\
		\includegraphics[width=0.5\textwidth]{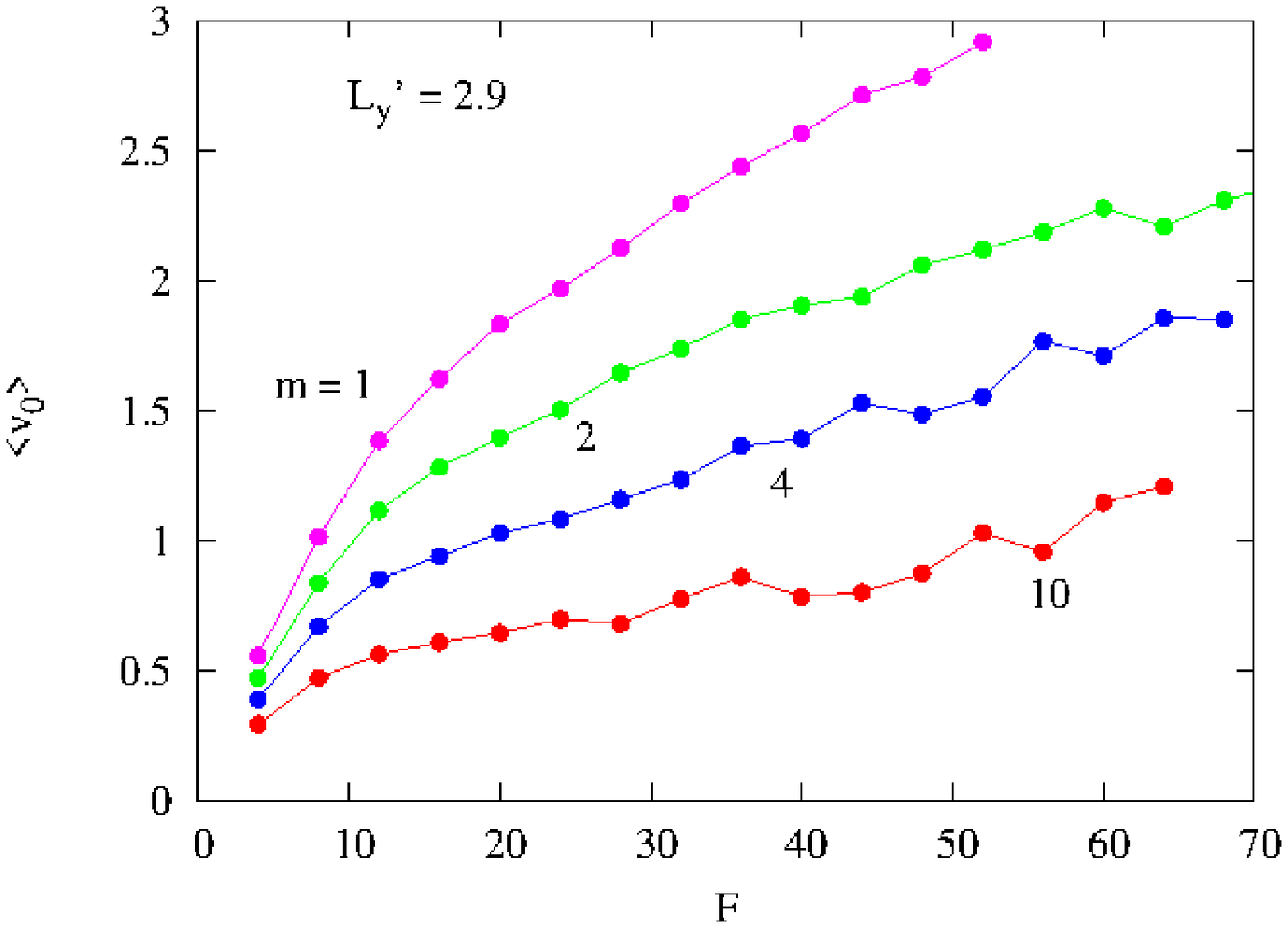} 
	\end{center}         
	\caption{\small   Dependence of the mean tracer velocity on the   
		driving force  $F$ for a channel width  $L_y' = 2.8$ (top panel) and
		 $2.9 $ (bottom panel). 	The masses of the neutral 
		particles differ from curve to curve  as indicated by the labels.}
	\label{NDM1}
\end{figure}
top and bottom panels of Fig \ref{NDM1} for $L_y' = 2.8$ and $2.9$, respectively.
For  $L_y'  = 2.8$, one observes  very strong negative mobility.
But an increase of the channel width by a moderate amount to  $2.9$ gives a totally
different picture. The slopes of the characteristic curves always remain positive.

\begin{figure}[!ht]
	\begin{center}
		\includegraphics[width=0.5\textwidth]{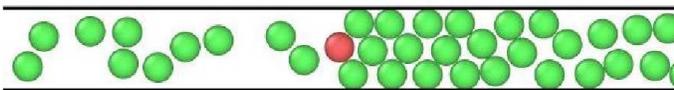} 
	\end{center}
	\caption{\small   Instantaneous configuration in a small  neighborhood of the tracer (red disk).
		The neutral particles (green disks)  accumulate in front of the tracer and 
		form crystal-like structures, which do not dissolve easily and slow down the  tracer.   
		 The channel width $L_y' = 2.7$. 
		The masses of the tracer and of the neutral particles  are set to unity,
		the driving force $F = 70$. }
	\label{crystal}
\end{figure}
The origin of this strange and unexpected behavior may be understood by inspecting a snapshot of the particle configuration 
in the neighborhood of the tracer. See Figure \ref{crystal}. The (green) neutral particles accumulate in front of the  tracer, which diffuses in the direction of $F$. The particles in front condense to  crystal-like structures, which do not dissolve easily and which become longer and more stable with increasing force $F$. In the limit of very large $F$ all neutral particles are included in this cluster, and the average speed of the tracer becomes minimal but still remains positive. The transition from the Ohmic regime for small $F$ to this nearly blocked transport for large $F$ is characterized by the negative differential mobility mentioned above. 

For a channel width larger than $2.8$, the tracer and the neutral particles have a tendency
to stay closer to one of the long channel boundaries than in the channel center. If this happens,  the tracer
has no difficulty to pass other neutral particles 
close to the other boundary which, consequently, enhances its speed
and disallows NDM. Also the very narrow channels geometrically do not favor the nucleation of 
crystal-like clusters in front of the tracer, which could block its advance. Consequently,  NDM is 
also not observed in this limit.

At this stage a few comments are in order: 

i) We have stressed above the importance of the boundary collisions for the nonlinear transport. 
 We have verified that NDM also takes place
- albeit for stronger forces $F$ - if the thermal walls are replaced by simple elastically-reflecting 
boundaries. Thus, this effect appears to be very robust. For small $F$, however, the nature of this boundary becomes unimportant~\cite{cividini_m_p2017}.

ii) In our quest for absolute negative mobility,  also  modifications of the  equations of motion (\ref{eom2}) were investigated. 
In one attempt,  an attractive  harmonic force acting on the $y$ coordinate of the tracer was added.
 This increases its position probability in the center of the channel.  In another, also a short-range force between
 the tracer and bath particles was added, which was designed to enhance the passing probability
 of these particles when close by.  None of these attempts have indicated ANM. 
Unfortunately,  the parameter space increases enormously by such attempts  and has not yet been
 explored to any significant extent.

\section{Conclusion}
\label{section:ccl}

In this work we exhibited two instances of negative response. The first one features ANM around an equilibrium state and occurs in a simple lattice model. We are able to predict this phenomenon analytically for small biases and shed light on the underlying mechanism in very dense systems. Keeping the same spirit, we then probed a more realistic system composed of hard disks in a channel. No ANM was found in that case, however we showed that a 'crystallization' of the neutral Brownian particles  at large biases can lead to NDM.

It is natural to ask which kind of modification would allow ANM in the hard-disks model in a way similar to the lattice model. Since the drive is homogeneous, the model as studied in this work cannot be expected to exhibit ANM in linear response around equilibrium. However, in principle it could exhibit ANM at large drive, something which we have not observed. Extending the model, \textit{e.g.} by introducing a $y$-dependent force or an extra potential, would introduce a second driving field that could reproduce the behaviour observed in the lattice model. 
Emulating the exchange move in the hard-disks system is however not such an easy task. Other variants also come to mind, like changing the masses or the radii of the particles, not to mention their shapes. The parameter space being extremely large, we leave this question open for future research.

\section*{Acknowledgments}

We thank M.~Aizenmann, H.~van Beijeren, B.~Derrida, Y.~Kafri, A.~Kundu, S.N.~Majumdar and A.~Miron for interesting discussions about this problem. One of us (HAP) wants to acknowledge the hospitality and support of the Weizmann Institute of Science, where parts of the work reported 
here have been performed. The support of the Israel Science Foundation (ISF) is gratefully acknowledged. The molecular dynamics simulations were carried out on the Vienna Scientific Cluster (VSC).
We are grateful for the generous allocation of computer resources. 


\appendix 

\section{$\alpha$ and $\gamma_+$ coefficients}
\label{section:appagag}

Here we give the coefficients that appear in the density profile calculation of Section~\ref{section:linres} and a few limits. Their expressions are
\begin{widetext}
\begin{eqnarray}
\label{eq:ag}
\alpha &=& \frac{1}{\Xi} \frac{(1-\rhob) \rhob}{2 (1-X)} \left(1+r(1-\rhob)+4 \rp \rhob (1-\rhob) + 2 \rp (1-\rhob) X (2 \rhob - 1)\right), \nonumber \\
\gamma_+ &=& \frac{1}{\Xi} \frac{\rp (1-\rhob)^2 \rhob (1-2 \rhob)}{X (1-X)}, \nonumber \\
\alpha' &=& \frac{1}{\Xi} \frac{(1-\rhob) \rhob }{2 (1-X)} \left(-r (1-\rhob) + (2 \rhob-1)(1+2\rp (1+X)(1-\rhob) \rhob)\right), \\
\gamma_+' &=& \frac{1}{\Xi} \frac{r (1-\rhob)^2 \rhob^2}{X (1-X)}, \nonumber
\end{eqnarray}
with the shorthand
\begin{equation}
\label{eq:defxi}
\Xi = \rp^2 (1+X) (1-2\rhob)^2 (1-\rhob) \rhob +r \rhob +r^ 2 \rhob (1-\rhob) +\rp (1-2\rhob)^2 +r \rp (1-\rhob) (1+\rhob (-2+(3+X)\rhob)),
\end{equation}
and $X$ is defined in~\eqref{eq:Xdef}.
\end{widetext}

For low densities $\rhob \to 0$ we get
\begin{eqnarray}
\label{eq:lowrhocoeffs}
X &\sim& - \frac{\rp}{1+r} \rhob, \nonumber \\
\Xi &\sim& \rp (1+r), \\
\alpha \sim - \alpha' &\sim& \frac{\rhob}{2 \rp}, \nonumber \\
\gamma_+ \sim \frac{1}{r \rhob} \gamma_+' &\sim& - \frac{1}{\rp}. \nonumber 
\end{eqnarray}

In the high-density limit $\rhob \to 1$ we have
\begin{eqnarray}
\label{eq:highrhocoeffs}
X &\sim& - \rp (1-\rhob), \nonumber \\
\Xi &\sim& r + \rp, \\
\alpha \sim \alpha' &\sim& \frac{1-\rhob}{2 (r+\rp)}, \nonumber \\
\gamma_+ \sim - \frac{\rp}{r} \gamma_+' &\sim& \frac{1-\rhob}{r+\rp}. \nonumber 
\end{eqnarray}
The values of $\alpha$ and $\alpha'$ are consistent with the high-density calculation~\eqref{eq:denhighrho} and the terms involving $\gamma$ and $\gamma'$ are negligible in the high-density limit.

One can also simplify
\begin{eqnarray}
\label{eq:signC1}
\alpha'+\gamma_+' X &=& (2 \rhob - 1) \frac{\rhob (1-\rhob)}{2 \Xi (1-X)} \\ && \times (r(1-\rhob) + 1 + 2 \rp (1+X) (1-\rhob) \rhob), \nonumber
\end{eqnarray}
which shows that the sign of the exponential layer changes when $\rhob = \frac{1}{2}$ for $\delta = 0$.

\section{Bath particle current}
\label{section:appJb}

Here we prove expression~\eqref{eq:Jb} for the bath particle current. In the lab frame, let us consider a given link between two sites, say $0$ and $1$, and examine the processes leading to a hop of a bath particle across this link.
\begin{itemize}
	\item One possibility is that the tracer occupies site $0$, that site $1$ is vacant and that site $2$ is occupied by a bath particle. In that case an exchange can occur and the bath particle can cross the designated link from right to left. The required configuration occurs with probability $\frac{\langle (1-\tau_1) \tau_2 \rangle}{L}$, the transition happens with a rate $\pp$ leads to an algebraic current $-1$ across the link.
	\item A similar contribution is obtained from the case where the tracer occupies $-1$, $0$ is vacant and $1$ is occupied, giving another factor $-\pp \frac{\langle (1-\tau_1) \tau_2 \rangle}{L}$. Two symmetric processes happen with rate $\qp$ instead of $\pp$, giving $2 \qp \frac{\langle (1-\tau_{L-1}) \tau_{L-2} \rangle}{L}$.
	\item If $0$ is occupied and $1$ is vacant, the bath particle can hop with rate $1$. This can happen if, initially, the bath particle occupies one of the positions $l=1,\ldots,L-2$ in the tracer frame, each one with probability $\frac{1}{L}$. For each $l$ this gives a factor $\frac{\langle \tau_l (1-\tau_{l+1}) \rangle}{L}$.
	\item If $0$ is vacant and $1$ occupied, a symmetrical process takes place. This gives a factor $\frac{\langle \tau_l (1-\tau_{l-1}) \rangle}{L}$ for each $l=2,\ldots,L-1$.
\end{itemize}
Adding all the factors, we get
\begin{eqnarray}
\label{eq:Jbproof}
\Jb &=& \frac{1}{L} \langle -2 \qp (1-\tau_{1}) \tau_{2} + 2 \qp (1-\tau_{L-1}) \tau_{L-2} \rangle \nonumber \\
&& + \frac{1}{L} \langle \sum_{l=1}^{L-2} \tau_l (1-\tau_{l+1}) - \sum_{l=2}^{L-1} \tau_l (1-\tau_{l-1}) \rangle.
\end{eqnarray}
The sums on the second line cancel out except for the difference $\tau_1 - \tau_{L-1}$ and~\eqref{eq:Jbproof} eventually gives expression~\eqref{eq:Jb} after using the mean-field approximation.


\section{Positivity of $\muee$}
\label{section:appmuee}

Here we show that $\muee$ as obtained in~\eqref{eq:muee} is always strictly positive for $(\rhob, r,\rp) \in (0,1) \times (0,\infty) \times (0,\infty)$. The numerator obviously vanishes at $\rhob =0$, $\rhob = 1$ and $r =0$ and is strictly positive otherwise. We express the denominator as a function of $U = r (1-\rhob)$ and $V = \rp \rhob (1-\rhob)$,
\begin{equation}
\label{eq:denommuee}
D = (2 \rhob - 1) (\frac{U}{1-\rhob}+\frac{2 \rhob-1}{\rhob(1-\rhob)} V) \sqrt{1+U} + U \sqrt{1+U + 4 V},
\end{equation}
and we will show that $D > 0 $ for all $(\rhob, U,V) \in (0,1) \times (0,\infty) \times (0,\infty)$. For fixed $U$ and $V$, $D \to + \infty$ for both $\rhob \to 0$ and $\rhob \to 1$. The denominator $D$ reaches a minimum inside the interval, for 
\begin{equation}
\label{eq:rhomin}
\rhob_\text{min} = \frac{-V + \sqrt{U V + V^2}}{U}.
\end{equation}
The value of this minimum is
\begin{eqnarray}
\label{eq:Dmin}
\min_{\rhob \in (0,1)} D &=& U \left( \sqrt{1+U+4V} - \sqrt{1+U}\right) \nonumber \\ &&+ 2 \sqrt{V} \left( \sqrt{U+V} - \sqrt{V}\right),
\end{eqnarray}
which is clearly positive, hence the positivity of $\muee$.

\section {The van Beijeren thermostat}
\label{section:appvbthermostat}

For a thermal wall it is required that a colliding particle   exchanges energy with the wall such that  it  is reflected with a velocity 
corresponding to the equilibrium temperature $ T_b$ of the wall. 
In the simplest version only the normal component of the velocity,
$ v_n = \mathbf{v} \cdot \mathbf{n}(\mathbf{q})$, is mapped onto $v_n'$, and the parallel component $v_p$
remains unchanged~\cite{vanbeijeren2014, bosetti_p2017}. Here, $'$ refers to the outgoing particle, 
and $\mathbf{n}(\mathbf{q})$ is a unit vector at position $\mathbf{q}$ and normal to the wall,  pointing inward.  
Instead, we require in the  following  that the particle is reflected from the wall with an
outgoing angle as in a specular reflection, but with its speed $v \equiv |\mathbf{v}| $ mapped according to
$ v' = g(v) v $.  To determine $g(v)$, a detailed balance condition is imposed, which requires that the 
collision frequencies with impact velocities $\mathbf{v}$ and $ - \mathbf{v}'$ are equal~\cite{vanbeijeren2014}.  
For particles in equilibrium with the wall,  the number of particles reflected into $\mathbf{v}'$ is the
same as under specular reflection. The average number of collisions per unit wall area with 
velocity between $\mathbf{v}$ and $\mathbf{v} + d\mathbf{v}$ is given by \cite{vanbeijeren2014}
\begin{equation}
n_c(\mathbf{v},\mathbf{q}) d\mathbf{v} = - v_n f^{(1)}(\mathbf{v}.\mathbf{q}) d\mathbf{v},
\end{equation}
where $f^{(1)}(\mathbf{v},\mathbf{q}) $ is the one-particle distribution function, which in the
canonical ensemble becomes
\begin{equation}
f^{(1)}(\mathbf{v},\mathbf{q})  = n(\mathbf{q}) \left( \frac{m \beta}{2 \pi} \right)^{d/2} 
\exp \left( - \frac{\beta m v^2}{ 2 } \right).
\label{detailed}     
\end{equation}   
$ n(\mathbf{q})$ is the particle density at $  \mathbf{q} $.
As usual, $\beta = 1/k_B T_b$, $k_B$ is the Boltzmann constant, and $d$ is the dimension.
In the following we  consider a planar system, $d = 2$. Then
the detailed balance condition  becomes
\begin{equation}
-v_n\exp\left(-\frac{\beta m \mathbf{v}^2}2\right) d \mathbf{v} = v'_n\exp\left(-\frac{\beta m(\mathbf{v}')^2}2\right) d\mathbf{v}'.
\label{three}
\end{equation}
Using polar coordinates, $ d \mathbf{v} = v dv d\phi; \;\; v_n = v \cos{\phi},$ one finds
\begin{eqnarray}
&& v \cos(\phi) \exp\left(-\frac{\beta m \mathbf{v}^2}2\right) dv d\phi \\ &=&  v' \cos(\phi') \exp\left(-\frac{\beta m(\mathbf{v}')^2}2\right) dv' d \phi'. \nonumber 
\end{eqnarray}
The integral over the angles on both sides of this equation cancel each other. Using the dimensionless abbreviations
$ X = \sqrt{( m \beta / 2)} \;v$ and   $ X' = \sqrt{ (m \beta/2)} \;v'$, a final integration on both sides 
yields
\begin{equation}
X' e^{-X'^2} - \frac{\sqrt{\pi}}{2} \mbox{erf}(X') = \mbox{const} - X e^{-X^2} + \frac{\sqrt{\pi}}{2} \mbox{erf}(X),
\label{g} 
\end{equation}
where $\mbox{erf}$ denotes the error function.
The integration constant
$$\mbox{const} =  - \frac{\sqrt{\pi}}{2}$$
is determined from the requirement that $v \to 0 $ is mapped into $v' \to \infty$, and $v \to \infty$ into $v' \to 0$.
For given $X$, respective $v$, a numerical solution of Eq.(\ref{g}) yields the map $v' = g(v) v$. 
The new velocity components normal and parallel to the wall after the collision become
\begin{equation}
v'_n = g(v) v_n; \;\;\; v'_p = - g(v) v_p.
\end{equation}

This map is time reversible and deterministic, $( - \mathbf{v}' )' = \mathbf{v}$. For large (small) velocities the energy transfer due to
a wall collision is negative (positive). Since this
condition prevails whenever the instantaneous kinetic temperature of the gas is larger  (smaller) than the wall temperature,
$T >   T_b$ (respective $T < T_b$), energy is transferred from the gas (wall) to the wall (gas) as required for a thermostat.     

A crucial test is the distribution function of the velocity component perpendicular to the wall, which in equilibrium
is expected to behave as~\cite{tehver1998}
\begin{equation}
\Phi(v_n) = m \beta | v_n| \exp \left( - \frac{ m \beta v_n^2}{2 } \right) 
\label{d2}
\end{equation}    
The experimentally obtained  distribution is in full agreement with
this theoretical prediction.  

%

\end{document}